\documentclass[10pt,aps,prl,twocolumn,showpacs,amsmath,amssymb,superscriptaddress,nofootinbib]{revtex4-1}
\usepackage{graphicx}
\usepackage{natbib}
\usepackage{amsmath}
\usepackage{xcolor}
\usepackage{comment}
\usepackage{soul}
\usepackage{xfrac}
\usepackage{mathtools}
\usepackage[colorlinks=true, urlcolor  = blue, citecolor = blue, linkcolor = blue]{hyperref}

\begin{document}

\title{Competition Between Exchange and Magnetostatic Energies in Domain Pattern Transfer from BaTiO$_3$(111) to Ni Thin Film.}

\author{K\'{e}vin J. A. Franke}
\affiliation{School of Physics and Astronomy, University of Leeds, Leeds LS2 9JT, United Kingdom}

\author{Colin Ophus}
\affiliation{National Center for Electron Microscopy, Molecular Foundry, Lawrence Berkeley National Laboratory, Berkeley, California 94720, USA}

\author{Andreas K. Schmid}
\affiliation{National Center for Electron Microscopy, Molecular Foundry, Lawrence Berkeley National Laboratory, Berkeley, California 94720, USA}

\author{Christopher H. Marrows}
\affiliation{School of Physics and Astronomy, University of Leeds, Leeds LS2 9JT, United Kingdom}

\date{\today}

\begin{abstract}
We use spin polarized low energy electron microscopy to investigate domain pattern transfer in a multiferroic heterostructure consisting of a $(111)$-oriented BaTiO$_{\mathrm{3}}$ substrate and an epitaxial Ni film. After in-situ thin film deposition and annealing through the ferroelectric phase transition, interfacial strain transfer from ferroelastic domains in the substrate and inverse magnetostriction in the magnetic thin film introduce a uniaxial in-plane magnetic anisotropy that rotates by $60^{\circ}$ between alternating stripe regions. We show that two types of magnetic domain wall can be initialized in principle. Combining experimental results with micromagnetic simulations we show that a competition between the exchange and magnetostatic energies in these domain walls have a strong influence on the magnetic domain configuration.

\end{abstract}
\maketitle

Domain pattern transfer from $(001)$-oriented BaTiO$_{\mathrm{3}}$ substrates into ferromagnetic thin films via interfacial strain transfer and inverse magnetostriction has been investigated intensively \cite{lahtinen_pattern_2011, Lahtinen_alternating_2012, chopdekar_spatially_2012, streubel_strainmediated_2013, Shirahata2015, Ghidini2015}. The multiferroic heterostructures thus obtained have been used to demonstrate electric field induced magnetization switching  \cite{lahtinen_pattern_2011, Shirahata2015, Ghidini2015}, magnetic domain wall motion \cite{PhysRevX.5.011010, Diego_electric_2017}, and electric field control of spin waves \cite{Hamalainen_PRA_2017, Huajun2021}, thus holding the promise of low-power spintronic devices.

BaTiO$_{\mathrm{3}}$ is a perovskite ferroelectric. Below $120^{\circ}$C, the cubic parent phase transforms into a tetragonal phase with a lattice elongation along an edge ($\langle001\rangle$). This distortion is associated with a shift of the positively charged Ti$^{4+}$ ions and negatively charge O$^{-}$ ions in opposite directions, giving rise to an electric polarization parallel to the lattice elongation. Strain relaxation leads to the formation of ferroelastic/ferroelectric domains with a perpendicular orientation of the lattice elongation and associated polarization in adjacent domains  \cite{merz_domain_1954, damjanovic_ferroelectric_1998}.

Recently, we reported on domain pattern transfer from $(111)$-oriented BaTiO$_{\mathrm{3}}$ substrates into Co thin films, where novel $60^{\circ}$ and $120^{\circ}$ domain wall structures are observed \cite{Franke_Co_prep}. Here we investigate the imprinting of the ferroelectric domains of $(111)$-oriented BaTiO$_{\mathrm{3}}$ substrates into thin epitaxial Ni films through a Pd buffer layer \cite{12}. Crucially, Ni exhibits a lower saturation magnetization $M_{\mathrm{s}}$ than Co, resulting in a reduced importance of magnetostatic effects. Combining spin polarized low energy electron microscopy (SPLEEM) \cite{Grzelakowski1996, Rougemaille2010} with micromagnetic simulations we show that the competition between exchange and magnetostatic energies has a significant (film thickness dependent) effect on the orientation of the magnetization in the magnetic domains. 

\begin{figure}
\centering
\includegraphics{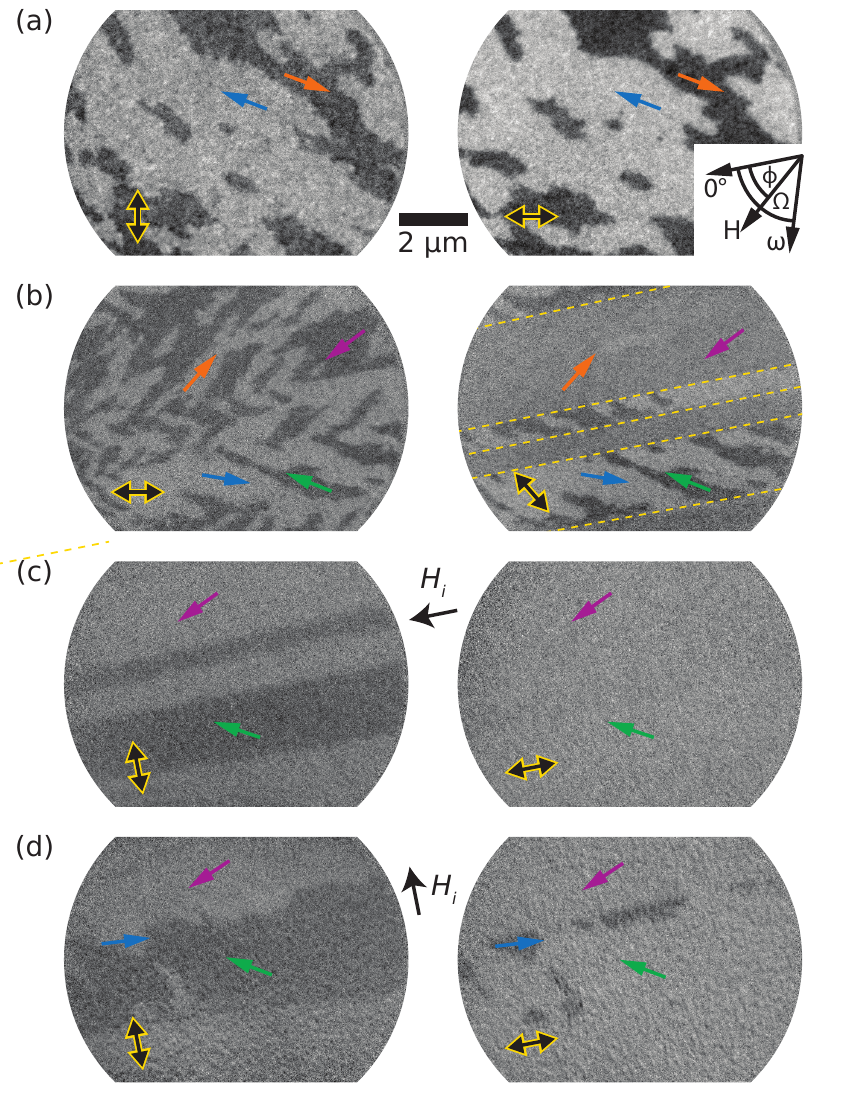}
\caption{\label{BTO_images} SPLEEM images of $3$ nm thick Ni film. Double-headed arrows indicate the axis of magnetic contrast and single-headed arrows the local orientation of the magnetization. (a) After deposition of the film. (b) After annealing the sample. After application of an initializing magnetic field $H_{i}$ parallel (c) and perpendicular (d) to the stripe domains.}
\end{figure}

\begin{figure}
\centering
\includegraphics{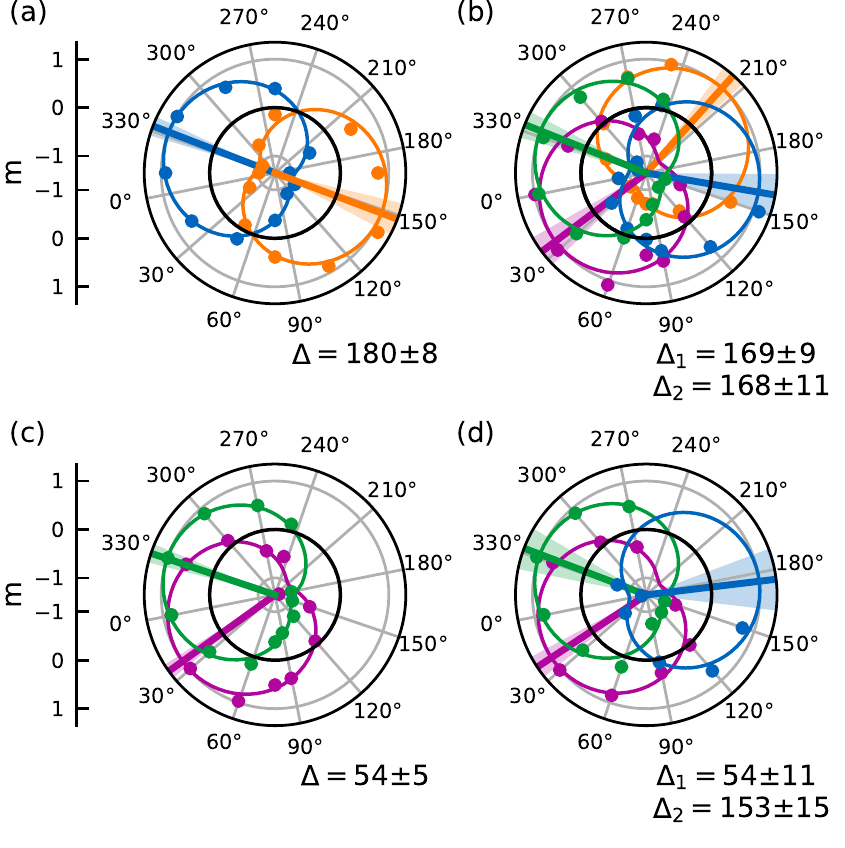}
\caption{\label{BTO_angular} Angular dependence of the normalized magnetic contrast corresponding to the images in Figure~\ref{BTO_images}. The data (dots) is compared to a fit (solid line). Radial lines indicate the resulting orientation of the magnetization with a transparent area indicating the uncertainty.}
\end{figure}

%Thin films are deposited in the SPLEEM instrument using molecular beam epitaxy. First, a Pd seed layer \textcolor{red}{Chris: how thick is the Pd?} is deposited onto the BaTiO$_{\mathrm{3}}$ substrate followed by the growth of a $3$ nm thin Ni layer, both at room temperature.  
%\textcolor{red}{Chris: both Pd and Ni are 111-oriented? can you say e.g that the 111 orientation was confirmed by some method.} 
The in-plane magnetization components of the as-deposited film are shown in Figure~\ref{BTO_images}(a) for two orthogonal directions of magnetic contrast (indicated with double-headed arrows). 
%\textcolor{red}{Chris: Do we know (or can we infer) the crystallographic directions at this point? }
%With assumption that images are 10 um wide (Gong told me that), each pixel is 21.6 nm
To determine the local magnetization direction, we take a series of images for various angles $\Omega$ of the in-plane contrast direction. The normalized contrast $m$ for the two largest domains as a function of $\Omega$ is shown as dots in Figure~\ref{BTO_angular}(a).
The data correspond to the projection of the magnetization onto the contrast direction and are therefore fitted to $m = \cos(\Omega-\phi)$. The magnetization angle $\phi$ thus obtained is indicated by a radial line. The shaded area of the same color indicates the uncertainty. The local magnetization direction is also plotted with arrows of the same color in Figure~\ref{BTO_images}.

The domain pattern we observe in the as-deposited film [Figure~\ref{BTO_images}(a)] is somewhat random, but the magnetization aligns along a uniaxial direction with the magnetization rotating by $180^{\circ}$ between neighboring domains. After heating the sample to $300^{\circ}$C, well into the cubic phase, and returning to room temperature, the domain pattern changes significantly [Figure~\ref{BTO_images}(b)]: An image taken with the magnetic contrast in the horizontal direction (left) reveals a zig-zag stripe pattern. An image taken with the contrast direction rotated by $49^{\circ}$ (right) helps clarify the origin of this pattern. The magnetization aligns along a uniaxial direction that rotates by $60^{\circ}$ between alternating stripe regions. Within these regions, a demagnetization pattern of domains separated by $180^{\circ}$ domain walls is observed.

\begin{figure}
\centering
\includegraphics{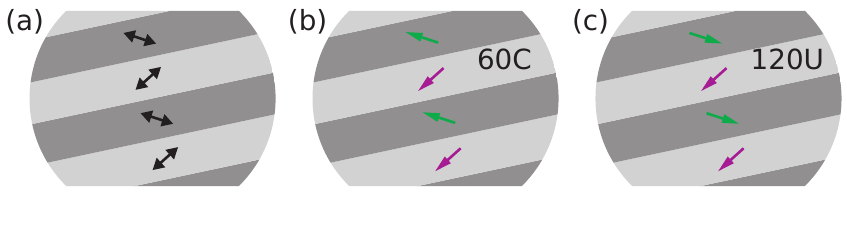}
\caption{\label{GeometrySketch} (a) Schematic illustration of the magnetic anisotropy pattern induced by strain coupling to the $(111)$-oriented ferroelectric BaTiO$_{\mathrm{3}}$ substrate. From this, two energetically distinct magnetization alignments are possible as sketched in (b) \& (c).}
\end{figure}
These images indicate the presence of a uniaxial anisotropy, the easy axis of which rotates by $60^{\circ}$ between adjacent stripe regions, therefore forming domains of magnetic anisotropy as sketched in Figure~\ref{GeometrySketch}(a). This spatial modulation of the in-plane anisotropy is a result of coupling to ferroelastic domains in the ferroelectric substrate. When projected onto the $(111)$-direction, the angle between the lattice elongations (discussed above) in adjacent ferroelastic domains is $60^{\circ}$ \cite{Franke_Co_prep}. Interfacial strain transfer from spatially modulated lattice elongations after heating the sample through the ferroelectric phase transition and inverse magnetostriction in the Ni film, yield the anisotropy pattern observed here.

We apply an initializing magnetic field $H_{i}$, saturating the sample and then returning to remanence, along the anisotropy domains. As a result, the demagnetization pattern is erased, and uniform magnetic domains are formed as shown in Figure~\ref{BTO_images}(c). The spin rotation (i.e. difference in magnetization angle $\phi$) between the two largest domains is $\Delta= 54 \pm 5^{\circ}$, close to the $60^{\circ}$ between anisotropy domains. We therefore conclude that the magnetization largely aligns with the anisotropy directions. It forms the domain pattern sketched in Figure~\ref{GeometrySketch}(b), which has been labeled ``$60$C'', as the magnetization rotates by $60^{\circ}$ between adjacent domains, and the domain wall is magnetically charged because of the accumulation of a net (virtual) magnetic charge due to the head-to-head (or tail-to-tail) magnetization configuration. Yet, significant deviations of the magnetization direction from the uniaxial anisotropy direction are observed. In particular the narrow dark domain in Figure~\ref{BTO_images}(c) exhibits a lower contrast with its neighboring domains than the wider domain of similar contrast below. 

\begin{figure}
\centering
\includegraphics{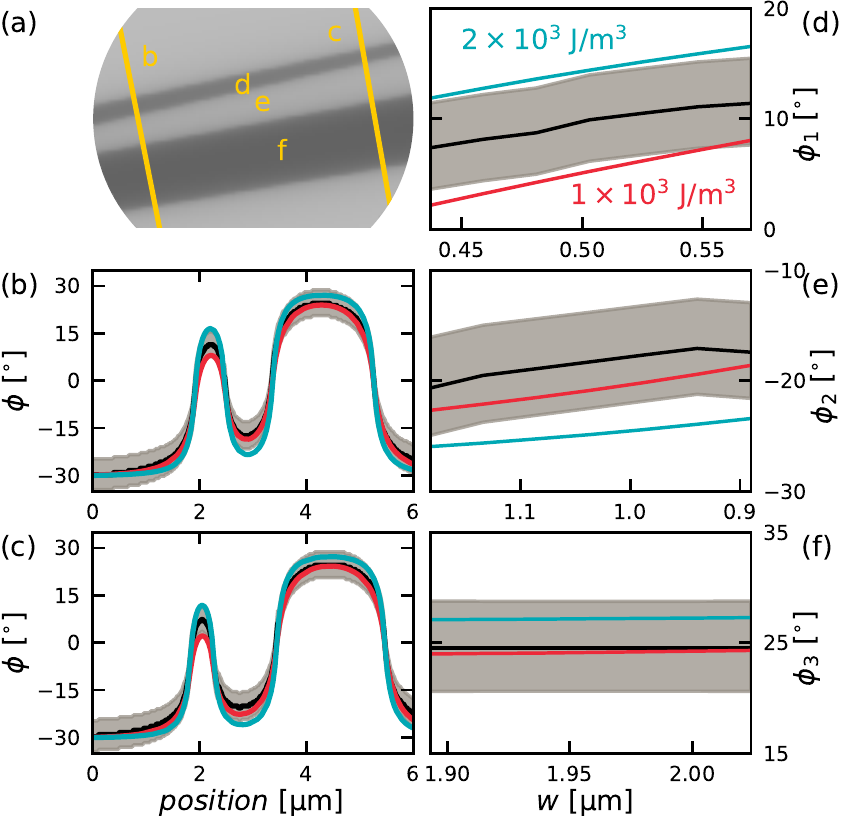}
\caption{\label{ani_fit}Determination of the anisotropy constant: (a) Fit to the image in Figure~\ref{BTO_images}(c). (b) \& (c) Profiles of the magnetization angle along the yellow lines in panel (a). The experimental data (black lines with grey shading indicating the uncertainty) is compared to micromagnetic simulations for a uniaxial anisotropy of $K_{u} =1 \times 10^{3}$ and $2 \times 10^{3}$ J/m$^{3}$ (red and turquoise lines). (d) -- (f) Magnetization angle $\phi$ at the center of the three central domains as a function of domain width $w$.}
\end{figure}

A fit of the image in Figure~\ref{BTO_images}(c) is shown in Figure~\ref{ani_fit}(a). We used nonlinear least squares to fit the spatial extent of the domains, using custom Matlab code. The fitting function $m$ for each boundary is given by the sigmoid function:
\begin{equation*}
m = \frac{\sfrac{r}{r_0}}{\left(1+\lvert \sfrac{r}{r_0} \rvert^{N}\right)^{\sfrac{1}{N}}}, 
\label{Kipswitch}
\end{equation*}
where $r_{0}$ sets the slope of the boundary at $r = 0$, and $N$ defines a power law which controls how quickly the boundary tails decay. Each boundary function was scaled to the best fit intensity levels of the domains on either side of each boundary. The best position $r_{\mathrm{boundary}}$ and rotation $\theta$ for all boundaries was fit simultaneously by parameterizing $(r = r' -r_{\mathrm{boundary}}) \times \cos(\theta).$

It can be observed that the domain walls are not perfectly parallel. This constitutes another indication that the magnetic domains are the result of strain coupling to ferroelectric domains in the BaTiO$_{\mathrm{3}}$ substrate, as the latter exhibits domains that are separated by ferroelectric domain walls that deviate from being perfectly parallel \cite{tagantsev2010domains}.

Figures~\ref{ani_fit}(b) \& (c) show the profiles of the magnetization angles (black lines, with gray shading indicating the uncertainty) taken along the location of the yellow lines in Panel (a). The profiles are subtly different, due to the width of domains not being constant. For example, in Panel (b), the domain labeled ``d'' is wider than in Panel (c), while the domains labeled ``e'' and ``f'' are both narrower. Panels (d) -- (f) show the magnetization direction at the center of these three domains as a function of domain width.  We do not observe a clear trend over this limited range of domains widths, but we compare the data in Figure~\ref{ani_fit}(b) -- (f) to micromagnetic simulations to estimate the magnitude of the uniaxial anisotropy.

We use OOMMF  \cite{donahue_oommf_1999} to run simulations with experimentally obtained parameters for the film thickness, domain widths, and anisotropy directions. We use literature values for the saturation magnetization $M_{\mathrm{s}} = 5.3\times10^5$ A/m and exchange stiffness $A = 8\times 10^{-12}$ \cite{Niitsu_2020, stohr_Book}.
The magnetization angles simulated for uniaxial anisotropy constants $K_{u} = 1 \times 10^{3}$ J/m$^3$ (red) and $K_{u} = 2 \times 10^{3}$ J/m$^3$ (turquoise) are compared to the experimental data in Figures~\ref{ani_fit}(b) -- (f). The experimental data mostly lie between the simulated data, and we conclude that the effective uniaxial anisotropy induced in the Ni film is  $K_{u} = 1.5 (5) \times 10^{3}$ J/m$^3$. This value is in line with results obtained for Ni films deposited at room temperature onto $(001)$-oriented BaTiO$_{\mathrm{3}}$ substrates \cite{streubel_strainmediated_2013}, and about an order of magnitude lower than the one obtained for Co on $(111)$-oriented BaTiO$_{\mathrm{3}}$ \cite{Franke_Co_prep}.

After applying the initializing magnetic field $H_{i}$ perpendicular to the anisotropy domains, one would expect the magnetization to align in the manner sketched in Figure~\ref{GeometrySketch}(c). That configuration is labeled ``$120$U'', as the magnetization rotates by $120^{\circ}$ between adjacent domains, and the head-to-tail orientation does not yield a net magnetic charge, thus making the domain walls uncharged. However, this configuration does not correspond to the experimental observation shown in Figure~\ref{BTO_images}(d) \cite{123}. We find that the 60C configuration is mostly initialized again, with a few regions [marked by a blue arrow in Figure~\ref{BTO_images}(d)] pointing in a direction perpendicular to the one the magnetic field was applied in.

This result is surprising for two reasons: first, the magnetization in one of the domains has switched by $180^{\circ}$ from the direction it was initialized in by the applied magnetic field; second, the charged configuration is associated with a cost in magnetostatic energy over the uncharged configuration. One would thus expect the uncharged configuration to be initialized in Figure~\ref{BTO_images}(d), unless the increased contribution of the exchange energy of $120^{\circ}$ domain walls over $60^{\circ}$ domain walls exceeds the difference in magnetostatic energy. Clearly, in this sample, the 60C magnetization configuration is energetically favorable over the 120U one. Furthermore, the occurrence of regions with almost reversed magnetization [blue arrow in Figure~\ref{BTO_images}(d)] at the boundary with domains, making an angle of $\approx 150^{\circ}$ with the magnetization in the domains, most likely reduces the magnetostatic energy.

The competition between exchange and magnetostatic energies is made more complex by the interaction between magnetic domain walls: the magnetization angle between neighboring domains is reduced as the domain width is decreased and the tails of domain walls overlap, as observed in Figure~\ref{ani_fit}. Previous articles have shown that charged and uncharged domain walls in multiferroic heterostructures exhibit different widths \cite{Franke_Co_prep, PhysRevB.85.094423}. For a given spin rotation $\Delta$, the width of an uncharged domain wall is $\delta_{u} \propto \sqrt{\sfrac{A}{K_{u}}}$, whereas for charged domain walls the width $\delta_{c}  \propto \sfrac{M_{s}^{2}t}{K_{u}}$ depends on the film thickness $t$ and the saturation magnetization of the material. Similarly, the energy of an uncharged domain wall $E_{u} \propto \sqrt{A K_{u}}$, while numerical simulations suggest that the energy $E_{c}$ of charged domain walls roughly scales with $M_{s}^{2}\,t$ \cite{hubert_schafer_BOOK, hubert_charged_1979,PhysRevLett.112.017201}. 

These expressions give an indication of how the domain wall widths and energies are expected to scale with the magnetic parameters and the film thickness. Crucially, we see that for a large saturation magnetization $M_{s}$ and film thickness $t$, magnetostatic effects lead to charged domain walls that are wider and higher in energy than uncharged domain walls. Conversely, here we investigate a film of small thickness $t$ with low saturation magnetization $M_{s}$, where the domain wall widths and energies are more comparable.

To compare the magnetization angles and energies of magnetization configurations in our sample we use micromagnetic simulations once again, varying film thickness and domain width.
\begin{figure}
\centering
\includegraphics{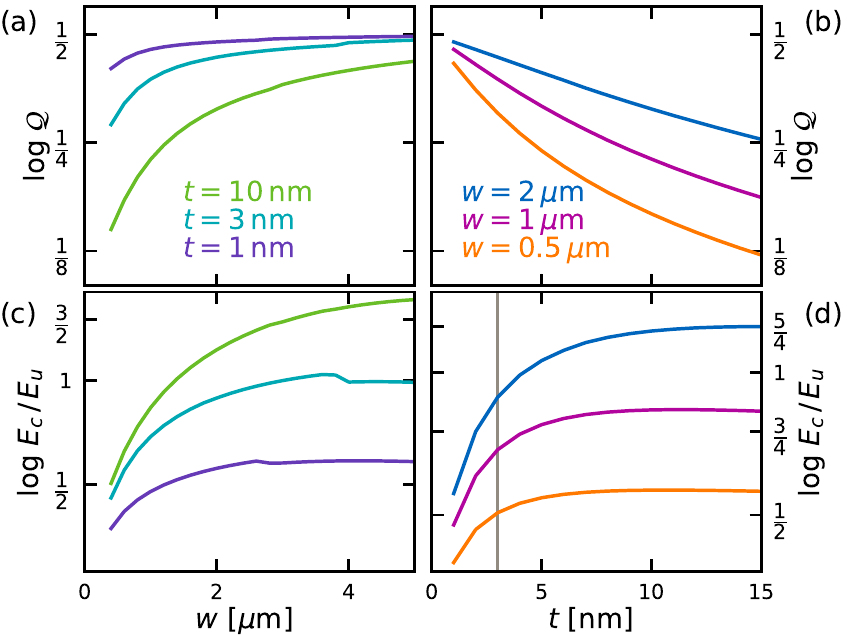}
\caption{\label{t-dep} Simulated dependence on (a) domain width, and (b) film thickness of the ratio $\mathcal{Q} = \Delta_{\mathrm{c}} / \Delta_{\mathrm{u}}$. Colors correspond to different film thicknesses or domain widths as indicated in the Figure. (c) \& (d) corresponding ratio between the energies of $60^{\circ}$ charged and $120^{\circ}$ uncharged configurations. The grey line indicates the $3$ nm film thickness of the sample.}
\end{figure}
Using the same parameters as above and two-dimensional periodic boundary conditions \cite{wang_twodimensional_2010} we simulate regular stripe domains of equal widths. Simulations are initialized with the magnetization tilted by $45^{\circ}$ from the stripe boundary towards the easy axes in the domains, such that the magnetization relaxes to the 60C or 120U configurations when the energy is minimized. %\textcolor{red}{Chris: I feel a bit more detail is needed here. How are the simulations constrained to produce these different wall types, for instance?} 

Figure~\ref{t-dep}(a) shows the ratio $\mathcal{Q} = \Delta_{\mathrm{c}} / \Delta_{\mathrm{u}}$ between spin rotations $\Delta$ of 60C and 120U configurations as a function of stripe width for three different film thicknesses. Similarly, Figure~\ref{t-dep}(b) shows the effect of film thickness on $\mathcal{Q}$. For isolated domain walls $\mathcal{Q}=\sfrac{60^{\circ}}{120^{\circ}}=\sfrac{1}{2}$. %\textcolor{red}{Chris: Is it exactly a half? seems a bit of a concidence. If not, might be better to write "a value close to 0.5"} 
We see that for thin films and wide stripes the data converge towards that value.

From the analytical expressions above we know that the widths of charged domain walls scale with the film thickness, while uncharged domain walls do not. As a result, domain wall tails of charged domain walls increasingly overlap when the film thickness is increased, while the width of charged domain walls remains unchanged. Generally, charged domain walls tend to be wider than uncharged ones, and thus overlap earlier when the domain width is reduced. Therefore $\mathcal{Q}$ is reduced when the film thickness is increased or the domain width decreased. 

Corresponding ratios between the energies of $60$C and $120$U configurations are shown in Figure~\ref{t-dep}(c) \& (d). We observe that for narrow domains in thin films the 60C configuration is energetically favorable, but the ratio between energies increases as the domains widen or the film thickness is increased. For very thin films ($1$ nm) or narrow domains, the low angle charged domain wall is always energetically favorable due to the lower exchange energy. For thick films ($10$ nm) and wide domains the difference in magnetostatic energy overcomes the difference in exchange energy, and the larger angle but uncharged configuration becomes energetically favorable, in line with previous observations \cite{Franke_Co_prep, PhysRevB.85.094423,PhysRevLett.112.017201}. We also see that for an intermediate thickness ($3$ nm) both domain walls are energetically equivalent, i.e. the differences in exchange and magnetostatic energies balance. By tuning the width of domains or adjusting the film thickness it is therefore possible to tune which configuration is the lower energy magnetization state in this heterostructure. 

To conclude, we used SPLEEM combined with micromagnetic simulations to investigate domain pattern transfer in a multiferroic heterostructure consisting of a $(111)$-oriented BaTiO$_{\mathrm{3}}$ substrate and an epitaxial Ni film. After in-situ thin film deposition and annealing through the ferroelectric phase transition, interfacial strain transfer from ferroelastic domains in the substrate and inverse magnetostriction in the magnetic thin film introduce a uniaxial in-plane magnetic anisotropy that rotates by $60^{\circ}$ between alternating stripe regions. Two magnetization configurations -- 60C and 120U -- can be initialized in principle. Our micromagnetic modelling shows that a competition between exchange and magnetostatic energies can be used to tailor the energy landscape of magnetic configurations and to choose which configuration is energetically favorable by tuning domain widths and film thickness. This explains the unexpected appearance of 60C walls after applying an initializing magnetic field $H_{i}$ perpendicular to the stripe regions in our experiments.  We expect these result to be transferable to other systems with spatial modulations of the anisotropy.

%Data associated with this publication will be made available from the University of Leeds repository \cite{Repository}.

\begin{acknowledgments}
This project has received funding from the European Union’s Horizon 2020 research and innovation programme under the Marie Sklodowska-Curie grant agreement No 750147. K.J.A.F. acknowledges support from the Jane and Aatos Erkko Foundation. Work at the Molecular Foundry was supported by the Office of Science, Office of Basic Energy Sciences, of the U.S. Department of Energy under Contract No. DE-AC02-05CH11231. This research used the Lawrencium computational cluster resource provided by the IT Division at the Lawrence Berkeley National Laboratory (Supported by the Director, Office of Science, Office of Basic Energy Sciences, of the U.S. Department of Energy under Contract No. DE-AC02-05CH11231.
\end{acknowledgments}

%\bibliography{references}

\begin{thebibliography}{26}
\bibitem [{12()}]{12}%
  \BibitemOpen
  \href@noop {} {}\bibinfo {note} {{Details of the in-situ thin film growth and
  imaging of magnetic contrast are described in reference
  \cite{Franke_Co_prep}.}}\BibitemShut {Stop}%
\bibitem [{123()}]{123}%
  \BibitemOpen
  \href@noop {} {}\bibinfo {note} {{Images in Figure~\ref{BTO_images}(d) were
  taken in a different location than the ones shown in
  Figure~\ref{BTO_images}(c) because the same location could not be found
  again. Still, both sets of images are representative for the domain patterns
  observed throughout the sample.}}\BibitemShut {Stop}%
\end{thebibliography}

%

\end{document}